\documentstyle[aps]{revtex}

\begin{document}
\title{Resonance fluorescence in a band gap material: Direct numerical simulation
of non-Markovian evolution}
\author{M. W. Jack$^{1,2}$ and J. J. Hope$^{1}$}
\address{$^{1}$Department of Physics, University of Auckland, \ \\
Private Bag 92019, Auckland, New Zealand\\
$^{2}$ Graduate School of Science and Engineering, Tokyo Institute of\\
Technology, Meguro-ku, Tokyo 152-8551, Japan}
\maketitle

\begin{abstract}
A numerical method of calculating the non-Markovian evolution of a driven
atom radiating into a structured continuum is developed. The formal solution
for the atomic reduced density matrix is written as a Markovian algorithm by
introducing a set of additional, virtual density matrices which follow, to
the level of approximation of the algorithm, all the possible trajectories
of the photons in the electromagnetic field. The technique is perturbative
in the sense that more virtual density matrices are required as the product
of the effective memory time and the effective coupling strength become
larger. The number of density matrices required is given by $3^{M}$ where $M$
is the number of timesteps per memory time. The technique is applied to the
problem of a driven two-level atom radiating close to a photonic band gap
and the steady-state correlation function of the atom is calculated.
\end{abstract}

\section{Introduction}

Photonic band-gap (PBG) materials give rise to an altered density of states
(DOS) for the electromagnetic field, which leads to a fundamentally
different atom-radiation field interaction \cite
{john87,ho90,joannopoulos97,yablonovich91,Gruning95,Gruning96,Ozbay94,Soukoulis96}%
. When atoms emit light into these structures, they exhibit inhibition of
spontaneous emission \cite{yablonovitch87,john94}, localization of light,
and the formation of an atom-photon bound state \cite{john84,john90,john95}.
These effects are due to entanglement between the atomic and photonic
states, and as such represent a failure of the Born-Markov approximation,
which allows the derivation of a master equation for the reduced density
matrix of the atom and has been an extraordinarily useful tool in the field
of theoretical quantum optics for treating spontaneous emission. The
spontaneous emission of photons into a PBG has been described analytically,
but describing such a system with the added nonlinearity of driving has
remained a theoretical challenge \cite{quang,molmer99}. This paper outlines
a numerical method that calculates the resonance fluorescence of a driven
atom in a PBG material. It may be applicable to other weakly non-Markovian
systems.

A photonic band gap is a range of frequencies in which no photon propagation
can occur. An atom near one of these band edges will experience an
interaction with an altered electromagnetic vacuum, and will therefore decay
in a different manner. Trivially, there will be no emission of photons into
the band gap, but the altered DOS near the band edge will have further
effects on the behavior. The case of spontaneous emission was treated by
John and Quang \cite{john94} who, due to the restricted state space in this
case, were able to solve the equations of motion for the system analytically
by Laplace transform methods. They found that the atom would not decay to
the ground state due to the presence of a localized atom-photon bound state
which was more strongly bound with stronger coupling. Near a band edge, the
energy of the propagating photons varies quadratically with the wavevector
in the same way as a massive particle propagating in free space. The model
of this system is therefore virtually identical to that of a single mode
atom laser \cite{hope97,moy97,moy99}.

The extension of the work of John and Quang to a driven atom has remained a
theoretical challenge. When pumping was added to the atom laser model, it
allowed access to a potentially infinite ladder of states within the lasing
mode. Fortunately, the linearity of the problem allowed a closed set of
equations for the relevant expectation values to be developed, and the
steady state output spectrum of the system was calculated \cite{hope00}. In
contrast, a driven two level atom in a PBG material has a much smaller
system (two states), but the nonlinear behavior of the atom due to the
driving, in conjuction with the output coupling to the field, does not allow
a closed set of equations to be deduced. Quang and John attempted a
treatment in which they assumed that no more than one photon was ever
present in the output field \cite{quang}, but in a subsequent comment this
assumption was shown to be equivalent to the Born approximation, and the
Monte Carlo wavefunction method employed to deal with this problem was shown
to be invalid \cite{molmer99}.

The memory of a system with a quadratic dispersion relation in one dimension
is effectively infinite. This is evident in the undriven case where photons
can be permanently localized near the atom, which can remain partially in
the excited state forever. Driving the atom in the presence of this dressed
atom-photon bound state will cause the bound state to become more highly
occupied. A driven atom in a PBG material is analogous to the pumped atom
laser model, which does not have a steady state without the inclusion of a
loss mechanism for the bound state \cite{Hope99}. In a real system the
build-up of energy in the bound state would reach some limit as
imperfections in both the PBG material and the quadratic approximation
became significant. In other words, describing the resonance fluorescence of
an atom in such a material requires inclusion of the losses in order to
produce a physical result. This will result in a finite memory time for the
evolution of the atomic system. In addition, the calculations so far have
dealt with an isotropic model of \ a bandgap but more appropriate for
artificially created bandgap materials is the anisotropic bandgap model \cite
{vats} which leads to memory function that decays much more rapidly than
that for an isotropic bandgap. In this paper we will present a method which
is able to calculate the evolution of a driven atom near a PBG providing the
memory time of the system is sufficiently short, and the coupling is
sufficiently weak.

In the next section we will describe our model for the driven atom, and
demonstrate how a formal solution for the reduced density matrix of the atom
can be generated. In Sec.~\ref{sec:method} we will describe a numerical
method for producing results with this formal solution. Sec.~\ref
{sec:results} shows how our method reproduces the exact result from a
similar, soluble system, and then describes the spectrum of the resonance
fluorescence near a band gap in a PBG material with varying levels of
detuning of the atomic resonance from the bandgap edge.

\section{Model}

\label{sec:model}

We consider a two level atom with atomic energy level separation $\hbar
\omega _{0}$. It is described by the Pauli spin matrices $\sigma $, $\sigma
^{\dagger }$ and $\sigma _{z}$, which satisfy the commutation relations $%
[\sigma ^{\dagger },\sigma ]=\sigma _{z}$, and $\{\sigma ^{\dagger },\sigma
\}=1$. The atom is driven by a classical field of frequency $\omega _{L}$,
and is modeled by the Hamiltonian 
\begin{equation}
H_{0}=\frac{\hbar \omega _{0}}{2}\;\sigma _{z}+\frac{\hbar \Omega }{2}%
\;\{\exp (-i\omega _{L}t)\;\sigma ^{\dagger }+\exp (i\omega _{L}t)\;\sigma
\},
\end{equation}
where $\Omega $ is the strength of the driving field experienced by the atom.

Under the rotating wave approximation \cite{louisell} the Hamiltonian for
the atom coupled to a large number of discrete modes of the electromagnetic
field can be written as 
\begin{equation}
H=H_{0}+\sum_{k}\left[ \hbar \omega _{k}b_{k}^{\dagger }b_{k}+i\hbar
(g_{k}^{\ast }b_{k}^{\dagger }\sigma -g_{k}b_{k}\sigma ^{\dagger })\right]
\end{equation}
where $b_{k}$ is the annihilation operator of the $k$-th mode, characterized
by the frequency $\omega _{k}$, and is coupled to the atom with the coupling
strength $g_{k}$. It is important to note that the coupling is linear in the
field operators.

For the formal manipulations which we will perform below it is convenient to
work in the interaction picture. In the interaction picture the Hamiltonian
can be written as 
\begin{equation}
H_{I}(t)=i\hbar \lbrack \xi ^{\dagger }(t)\sigma (t)-\xi (t)\sigma ^{\dagger
}(t)],
\end{equation}
where the {\em driving field}\cite{jack00}, $\xi (t)$, is defined by 
\begin{equation}
\xi (t)=\sum_{k}g_{k}b_{k}e^{-i\omega _{k}t},
\end{equation}
and the interaction picture atomic operator $\sigma (t)$ is given by $\sigma
(t)=U_{0}^{\dagger }(t)\;\sigma \;U_{0}(t)$. The free evolution operator $%
U_{0}$ in this equation is defined by $U_{0}(t)=e^{-\frac{i}{\hbar }H_{0}t}$.

A quantity fundamental to any attempt to describe this interaction in terms
of the atomic system alone is the memory function, defined as the commutator
of the driving field with its complex conjugate; 
\begin{equation}
f_{{\rm m}}(t,t^{\prime })=[\xi (t),\xi ^{\dagger }(t^{\prime })].
\label{eq:fdef}
\end{equation}
For $t>t^{\prime }$, the memory function is the probability {\em amplitude}
of absorbing a photon at time $t$ that was emitted previously at time $%
t^{\prime }$. It is therefore the Feynman propagator of a virtual photon.
Non-Markovian behavior arises when one cannot consider the lifetime of the
virtual photons to be vanishingly small. The Born-Markov approximation can
be made for free space resonance fluorescence, as (for sufficiently high
frequencies) the linear dispersion relation ($\omega _{k}=ck$) produces an
infinitely narrow memory function. Unfortunately, near a photonic band gap
the dispersion relations can be quadratic, and this leads to a very long
memory time.

The combination of non-Markovian evolution and the nonlinear behavior of a
driven atom renders this problem intractable to either Born-Markov master
equation methods applicable to free-space resonance fluorescence \cite{resfl}%
, or Laplace transform methods that have been applied to the case of
spontaneous emission into a band gap material \cite{john94}. Instead, we
approach this problem by a direct numerical solution of the evolution of the
reduced atomic system.

The first important step is the derivation of a formal equation for the
reduced density matrix of the atomic system. The reduced density matrix is
defined by tracing over the field degrees of freedom of the density matrix
for the full interacting system. In the interaction picture it has the form 
\begin{equation}
\rho (t)={\rm Tr}_{{\rm field}}\left\{ \overleftarrow{T}\left[ e^{-\frac{i}{
\hbar }\int_{0}^{t}dsH_{I}(s)}\right] \rho _{{\rm tot}}\overrightarrow{T} %
\left[ e^{\frac{i}{\hbar }\int_{0}^{t}duH_{I}(u)}\right] \right\} ,
\label{densitymatrix}
\end{equation}
where $(\overrightarrow{T})\overleftarrow{T}$ denotes (anti-)timeordering.
Assuming an initial factorized state of the atom and field, $\rho _{{\rm tot}
}=\rho \otimes \rho _{{\rm field}}$, and that the field is initially in a
generalized Gaussian state \cite{qn} (e.g. a thermal state), a formal
expression in terms of only system variables can be determined for the
right-hand side of Eq.(\ref{densitymatrix}). Here we will perform a
derivation for the case of an initial vacuum field, $\rho _{{\rm field}
}=|\{0\}\rangle \langle \{0\}|$.

We use the identity \cite{jack00} 
\begin{equation}
\overleftarrow{T}\left[ e^{-\frac{i}{\hbar }\int_{0}^{t}dsH_{I}(s)}\right] =%
\overleftarrow{T}_{\sigma }\left\{ V_{+}V_{0}V_{-}\right\}  \label{eq:udef}
\end{equation}
where $\overleftarrow{T}_{\sigma }$ is a time ordering operator which
denotes a time ordering on the atomic operators only, and where we have
defined the operators 
\begin{eqnarray}
V_{+} &=&\exp \left( -\int_{0}^{t}ds\;\xi ^{\dag }(s)\sigma (s)\right)
\label{eq:v+def} \\
V_{0} &=&\exp \left( -\int_{0}^{t}du\;\int_{0}^{u}dv\;f_{{\rm m}%
}(u,v)\;\sigma ^{\dag }(u)\sigma (v)\right) \\
V_{-} &=&\exp \left( \int_{0}^{t}ds\;\xi (s)\sigma ^{\dag }(s)\right)
\label{eq:v-def}
\end{eqnarray}
where $f_{{\rm m}}(u,v)$ is the memory function, defined by Eq.~(\ref
{eq:fdef}). Similarly, we can write 
\begin{equation}
\overrightarrow{T}\left[ e^{\frac{i}{\hbar }\int_{0}^{t}dsH_{I}(s)}\right] =%
\overrightarrow{T}_{\sigma }\left\{ V_{-}^{\dagger }V_{0}^{\dagger
}V_{+}^{\dagger }\right\}
\end{equation}
for the anti-timeordered term.

In order to greatly simplify the manipulation of these time-ordered
quantities we introduce a notational apparatus, first used by Feynman \cite
{feynman}, whereby unprimed atomic operators are assumed to be time-ordered
to the left of $\rho $ and primed operators anti-timeordered to the right.
We can then write Eq.~(\ref{densitymatrix}) as 
\begin{equation}
\rho (t)={\rm Tr}_{{\rm field}}\left\{ V_{+}V_{0}V_{-}(\rho
\otimes|\{0\}\rangle \langle \{0\}|) V_{-}{^{\prime }}^{\dagger }V_{0}{%
^{\prime }}^{\dagger }V_{+}{\ ^{\prime }}^{\dagger }\right\} ,
\label{primenotation}
\end{equation}
where $V_{j}{^{\prime }}=V_{j}(\xi ,\xi ^{\dag },\sigma {^{\prime },\sigma
^{\prime \dag }).}$ This notation allows us to manipulate the atomic
operators as if they were $c$-numbers while keeping track of their correct
ordering in the expression. We can use the cyclic properties of the trace
over the field variables to write Eq.(\ref{primenotation}) in the form 
\begin{equation}
\rho (t)=\langle \{0\}|V_{-}{^{\prime }}^{\dagger }V_{0}{^{\prime }}%
^{\dagger }V_{+}{^{\prime }}^{\dagger }V_{+}V_{0}V_{-}|\{0\}\rangle \rho .
\end{equation}
We now normally order the field operators in this expression by making use
of the relations $e^{A+B}=e^{A}e^{B}e^{-\frac{1}{2}[A,B]}=e^{B}e^{A}e^{\frac{%
1}{2}[A,B]}$ for two operators $A$ and $B$ satisfying $[A,[A,B]]=[B,[A,B]]=0$%
. The normally-ordered field operators are annihilated on the field vacuum
and we are left with the result 
\begin{equation}
\rho (t)=e^{{}{\cal L}{}(t)}\rho ,  \label{formalsol}
\end{equation}
where 
\begin{equation}
{\cal L}{}(t)=\int_{0}^{t}du\int_{0}^{u}dv\,\left\{ [{\sigma ^{\prime }}%
^{\dagger }(u)-\sigma ^{\dagger }(u)]f_{{\rm m}}(u,v)\sigma (v)+\left[
\sigma (u)-\sigma ^{\prime }(u)\right] f_{{\rm m}}^{\ast }(u,v){\sigma
^{\prime }}^{\dagger }(v)\right\} .
\end{equation}
This is a formal solution that contains no reference to the field operators.
The only reference to the particular characteristics of the field is via the
memory function. The $\sigma ^{\dagger }(u)\sigma (v)$ and $\sigma ^{\prime
}(u){\sigma ^{\prime }}^{\dagger }(v)$ terms arise from the vacuum field to
vacuum field evolution while the terms linking the two sides of the density
matrix, $\sigma ^{\prime \dagger }(u)\sigma (v)$ and $\sigma (u){\sigma
^{\prime }}^{\dagger }(v),$ arise from tracing over all the photons that are
irretrievably emitted into the field. Note that only under the Born-Markov
approximation [where it is possible to replace $f_{{\rm m}}(t,s)$ by $\gamma
\delta (t-s)$] does the evolution matrix, $\ e^{{}{\cal L}(t)},$ factorize
at each time, as in this case the double integral reduces to a single
integral. It is then possible to write down a ordinary differential equation
for $\rho (t)$. In general, however, this is not possible and it is
necessary to use the formal solution Eq.(\ref{formalsol}).

The theoretical difficulty with using this formal solution is not the size
of the system, which can be described by three real numbers, but with the
implicit time ordering of the system operators, which in general leads to an
infinite number of \ ordered terms. Physically, these higher order terms are
the processes associated with increasing numbers of photons, and will become
increasingly less significant in a system with a finite memory time. A
driven system without a finite memory time may lead to unphysical behavior 
\cite{hope00}, so we are particularly interested in systems with a finite
memory time. When this timescale is shorter than other dynamic timescales we
are describing a Markovian system, and otherwise we are describing a
non-Markovian system.

In this work we consider a numerical approach to solving the evolution
matrix $e^{{}{\cal L}(t)}$. We discretize time and then make the
approximation of a finite memory time $T_{{\rm m}}$. In other words, we
assume that we can effectively put $f_{{\rm m}}(t,s)=0$ for all $s<t-T_{{\rm %
m}}$. Our algorithm considers all possible processes that can occur with
these restrictions. From Eq.(\ref{formalsol}) it is clear that it is not
possible to determine $\rho (t+\Delta t)$ from only $\rho (t)$ as at time $t$
there are photons in the field, represented by the presence of the memory
function that will re-interact with the atom at time $t+\Delta t$. These
photons have to be taken into account in any algorithm. It turns out that
under the finite memory time approximation we can follow the evolution of
these photons by introducing a finite number of additional, ``virtual'',
density matrices, $\rho ^{k},$ $k\neq 0$. The evolution equation for the
reduced density matrix together with the virtual density matrices has the
form 
\begin{equation}
\rho ^{k}(t+\Delta t)=\sum_{j}D_{kj}(t)\rho ^{j}(t)  \label{tool}
\end{equation}
where $D_{kj}(t)=D_{kj}\left[ \sigma (t),\sigma ^{\dagger }(t),\sigma
^{\prime }(t),{\sigma ^{\prime }}^{\dagger }(t)\right] $ is an evolution
superoperator and $\rho ^{0}(t+\Delta t)\approx \rho (t+\Delta t),$
approximates the reduced density matrix.

The method is based on transforming the numerical algorithm for propagating
a non-Markovian differential equation into a Markovian form. To motivate
this idea, consider the following elementary example of an
integro-differential equation,

\[
\frac{dc(t)}{dt}=-i\omega _{0}c(t)-\int_{t-T_{{\rm m}}}^{t}f_{{\rm m}%
}(t-s)c(s). 
\]
with a finite lower-bound on the integral. This integro-differential
equation can be propagated numerically via the algorithm

\begin{equation}
c_{n+1}=(1-i\omega _{0}\Delta t)c_{n}-\frac{\Delta t^{2}}{2}\left[
f_{0}c_{n}+2\sum_{j=n-M+2}^{n}f_{n-j+1}c_{j}+f_{M}c_{n-M+1}\right] +O(\Delta
t^{2})  \label{routine}
\end{equation}
where $c_{n}=c(n\Delta t),f_{n}=f_{{\rm m}}(n\Delta t)$ and $M+1=T_{{\rm m}%
}/\Delta t.$ We have used the trapezoidal rule for the integral and Euler's
method to propagate the differential equation forward. Due to the finite
memory time, \ this algorithm can be written in an explicitly Markovian form
by introducing $M-1$ extra variables $b^{k}$. The equivalent Markovian
algorithm is

\begin{eqnarray*}
c_{n+1} &=&(1-i\omega \Delta t)c_{n}-\frac{\Delta t^{2}}{2}%
f_{0}c_{n}+b_{n}^{1}+O(\Delta t^{2}) \\
b_{n+1}^{1} &=&b_{n}^{2}-\Delta t^{2}f_{2}c_{n} \\
b_{n+1}^{2} &=&b_{n}^{3}-\Delta t^{2}f_{3}c_{n} \\
&&\vdots \\
b_{n+1}^{M-1} &=&-\frac{\Delta t^{2}}{2}f_{M}c_{n}
\end{eqnarray*}
with the initial condition $b_{0}^{k}=0.$ In this particular case, this
transformation does not offer any advantage, however, this example is
analogous to the algorithm for propagating the reduced density matrix that
we will present in the next section. In that case, due to the presence of
time-ordering, reformulation as a Markovian algorithm becomes invaluable.

\section{Numerical Method}

\label{sec:method}

As a first step in determining an algorithm let us evaluate the integrals in
the exponential in Eq.(\ref{formalsol}) on a grid, 
\begin{equation}
{\cal L}(t)=\sum_{j=0}^{j=N}\Delta tL_{j}+O(\Delta t^{2})
\end{equation}
where 
\begin{equation}
L_{j}=\sum_{k={\rm max}(j-M+1,0)}^{j}\!\!\!\!\!\!\!\!\!\Delta t\left\{ ({%
\sigma ^{\prime }}_{j}^{\dagger }-\sigma _{j}^{\dagger })f_{j-k}\sigma
_{k}+(\sigma _{j}-\sigma _{j}^{\prime })f_{j-k}^{\ast }{\sigma ^{\prime }}%
_{k}^{\dagger }\right\} .
\end{equation}
For convenience of notation we have written $t_{j}=j\Delta t$ where $t_{N}=t$%
, $a_{j}=a(t_{j})$ and $f_{n}=W_{n}f(n\Delta t).$ The weights are given by 
\begin{equation}
W_{n}=\left\{ 
\begin{array}{cc}
\frac{1}{2},\quad & n=0 \\ 
\frac{1}{2},\quad & n=M-1 \\ 
1,\quad & {\rm otherwise}
\end{array}
\right. ,
\end{equation}
as we are using the trapezoidal rule (as the time ordering produces an
integrand that is effectively discontinuous at each time step there is good
reason to believe that a higher order rule will not improve accuracy). Note
that we have also explicitly truncated the sums after $M\geq T_{{\rm m}%
}/\Delta t$ as the memory function is assumed to be zero for all additional
terms. We can now write 
\begin{eqnarray}
e^{{}{}{\cal L}(t)} &=&\prod_{j}e^{\Delta tL_{j}}+O(\Delta t^{2}),  \nonumber
\\
&=&\prod_{j}(1+\Delta tL_{j})+O(\Delta t^{2}),
\end{eqnarray}
where in the second line we have expanded out the exponentials at each time, 
$t_{j}$, to first order in $\Delta t$. Note that it is still necessary to
timeorder the operators in this expression. Let us attempt to write this
expression in terms of the above mentioned density matrices $\rho ^{k}$ at
one particular time $n-1$, where we choose $N-M>n>M$ as this is the most
common case. First split the evolution matrix at $n+M$ and $n$, 
\begin{equation}
e^{{}{\cal L}(t)}=\prod_{k=n+M}^{N}(\cdots )\prod_{j=n}^{n+M-1}(1+\Delta
tL_{j})\bar{\rho}+O(\Delta t^{2}),
\end{equation}
where it is convenient to separate out the factor $\bar{\rho}%
=\prod_{j=0}^{n-1}(1+\Delta tL_{j})\rho (0)$ as it does not take part in our
subsequent manipulations. Then split the sums over $k$ in $L_{j}$ for $j<n+M$
into parts containing interaction picture operators before and after $n$, 
\begin{equation}
(1+\Delta tL_{j})=x_{j,n}+\Delta t({\sigma ^{\prime }}_{j}^{\dagger }-\sigma
_{j}^{\dagger })y_{j,n}+\Delta t(\sigma _{j}-\sigma _{j}^{\prime })z_{j,n},
\label{sam}
\end{equation}
where 
\begin{eqnarray}
x_{j,n} &=&1+\Delta t\sum_{k=n}^{j}\Delta t\left[ ({\sigma ^{\prime }}%
_{j}^{\dagger }-\sigma _{j}^{\dagger })f_{j-k}\sigma _{k}+(\sigma
_{j}-\sigma _{j}^{\prime })f_{j-k}^{\ast }{\sigma ^{\prime }}_{k}^{\dagger }%
\right] \\
y_{j,n} &=&\sum_{k=j-M+1}^{n-1}\Delta tf_{j-k}\sigma _{k} \\
z_{j,n} &=&\sum_{k=j-M+1}^{n-1}\Delta tf_{j-k}^{\ast }{\sigma ^{\prime }}%
_{k}^{\dagger }.
\end{eqnarray}
Consider the $M$ factors from $j=n$ to $j=n+M-1$. Due to the truncation of
the sums at $M$ terms, it is only these factors that contain interaction
picture operators at time $n$. We can expand this {\em product} of $M$
factors, each of which has three terms, as a {\em sum} over $3^{M}$ terms 
\begin{equation}
\prod_{j=n}^{n+M-1}(1+\Delta tL_{j})=\sum_{m=0}^{3^{M-1}-1}g_{m,n}h_{m,n},
\label{peter}
\end{equation}
where $g_{m,n}$ contains only operators after $n-1$ and $h_{m,n}$ contains
only operators before $n$. Note that since $y_{n+M-1,n}=z_{n+M-1,n}=0$ we
only need to consider a sum over $3^{M-1}$ terms. To write down the explicit
form for $g_{m,n}$ and $h_{m,n}$ it is convenient to introduce a new
notation to label the terms. We write the index $m$ in trinary (i.e. base
three), where the $k^{{\rm th}}$ trit (bit in base three) tells us which of
the three factors to choose from the $(1+\Delta tL_{k})$ term (shown in Eq.~(%
\ref{sam}). Our new index then lists the trits which correspond to the first
term, $\{A_{1},\cdots ,A_{p}\}$; the second term, $\{B_{1},\cdots ,B_{q}\}$;
or the third term, $\{C_{1},\cdots ,C_{M-q-p}\}$. These lists replace the $m$
index. In this notation we can write 
\begin{eqnarray}
g_{n}^{\{A_{1},\cdots ,A_{p};B_{1},\cdots ,B_{q};C_{1},\cdots ,C_{M-q-p}\}}
&=&x_{n+M-1,n}\prod_{j=1}^{p}x_{n+A_{j}-1,n}\prod_{k=1}^{q}\Delta t({\sigma
^{\prime }}_{n+B_{k}-1}^{\dagger }-\sigma _{n+B_{k}-1}^{\dagger })  \nonumber
\\
&&\times \prod_{l=1}^{M-q-p}\Delta t(\sigma _{n+C_{l}-1}-\sigma
_{n+C_{l}-1}^{\prime })
\end{eqnarray}
and 
\begin{equation}
h_{n}^{\{B_{1},\cdots ,B_{q};C_{1},\cdots
,C_{M-q-p}\}}=\prod_{k=1}^{q}y_{n+B_{k}-1,n}%
\prod_{l=1}^{M-q-p}z_{n+C_{l}-1,n}.
\end{equation}
where, if the lists $\{B_{1},\cdots ,B_{q};C_{1},\cdots ,C_{M-q-p}\}$ are
the empty set, $\emptyset $, by construction we have $h_{n}^{\emptyset }=1$.
This allows us to define a set of $3^{M-1}$ density matrices at time $n-1$
of the form 
\begin{equation}
\rho _{n-1}^{\{B_{1},\cdots ,B_{q};C_{1},\cdots
,C_{M-q-p}\}}=h_{n}^{\{B_{1},\cdots ,B_{q};C_{1},\cdots ,C_{M-q-p}\}}\bar{%
\rho},
\end{equation}
Note that $\rho _{n-1}^{\emptyset }$ is defined such that it corresponds to $%
\rho (t_{n-1})$ up to order $\Delta t^{2}$.

If we can determine the transformation from $\rho _{n-1}^{k}$ to $\rho
_{n}^{j}$ we will have achieved our aim of writing the non-Markovian
evolution in terms of a Markovian evolution in an extended space. This is
done by writing $\rho _{n}^{j}$ in terms of $\rho _{n-1}^{k}$. Noting that $%
y_{j,n+1}=y_{j,n}+\Delta tf_{j-n}\sigma _{n}$ and $z_{j,n+1}=z_{j,n}+\Delta
tf_{j-n}^{\ast }{\sigma ^{\prime }}_{n}^{\dagger }$ we can write 
\begin{eqnarray}
\rho _{n}^{\{B_{1},...B_{q};C_{1},...C_{M-q-p}\}}
&=&\prod_{k=1}^{q}[y_{n+B_{k},n}+\Delta tf_{B_{k}}\sigma
_{n}]\prod_{j=1}^{M-q-p}[z_{n+C_{j},n}+\Delta tf_{C_{j}}^{\ast }{\sigma
^{\prime }}_{n}^{\dagger }]  \nonumber \\
&&\quad \times \lbrack x_{n,n}+\Delta t({\sigma ^{\prime }}_{n}^{\dagger
}-\sigma _{n}^{\dagger })y_{n,n}+\Delta t(\sigma _{n}-\sigma _{n}^{\prime
})z_{n,n}]\bar{\rho}.
\end{eqnarray}
Expanding out the first two sets of factors and keeping only first-order
terms the right hand side can now be written in terms of the density
matrices at $n-1$, 
\begin{eqnarray}
\rho _{n}^{\{B_{1},\cdots ,B_{q};C_{1},\cdots ,C_{M-q-p}\}} &=&\left[
1+\Delta t^{2}({\sigma _{n}^{\prime }}^{\dagger }-\sigma _{n}^{\dagger
})f_{0}\sigma _{n}+\Delta t^{2}(\sigma _{n}-\sigma _{n}^{\prime
})f_{0}^{\ast }{\sigma _{n}^{\prime }}^{\dagger }\right] \rho
_{n-1}^{\{B_{1}+1,\cdots ,B_{q}+1;C_{1}+1,\cdots ,C_{M-q-p}+1\}}  \nonumber
\\
&&+\Delta t\sigma _{n}\sum_{k=1}^{q}f_{B_{k}}\rho _{n-1}^{\{B_{1}+1,\cdots
,B_{k-1}+1,B_{k+1}+1,\cdots ,B_{q}+1;C_{1}+1,\cdots ,C_{M-q-p}+1\}} 
\nonumber \\
&&+\Delta t{\sigma ^{\prime }}_{n}^{\dagger
}\sum_{k=1}^{M-q-p}f_{C_{k}}^{\ast }\rho _{n-1}^{\{B_{1}+1,\cdots
,B_{q}+1;C_{1}+1,\cdots ,C_{k-1}+1,C_{k+1}+1,\cdots ,C_{M-q-p}+1\}} 
\nonumber \\
&&+\Delta t({\sigma ^{\prime }}_{n}^{\dagger }-\sigma _{n}^{\dagger })\rho
_{n-1}^{\{1,B_{1}+1,\cdots ,B_{q}+1;C_{1}+1,\cdots ,C_{M-q-p}+1\}}  \nonumber
\\
&&+\Delta t(\sigma _{n}-\sigma _{n}^{\prime })\rho _{n-1}^{\{B_{1}+1,\cdots
,B_{q}+1;1,C_{1}+1,\cdots ,C_{M-q-p}+1\}}+O(\Delta t^{2}).  \label{final}
\end{eqnarray}
Physically, during a short enough interval only one photon can be emitted or
absorbed at a time. The higher order terms in $\Delta t$ correspond to two
or more particle emissions or absorptions during this interval and hence we
have dropped them. That is, except for the terms $\Delta t^{2}({\sigma
_{n}^{\prime }}^{\dagger }-\sigma _{n}^{\dagger })f_{0}\sigma _{n}+\Delta
t^{2}(\sigma _{n}-\sigma _{n}^{\prime })f_{0}^{\ast }{\sigma _{n}^{\prime }}%
^{\dagger },$ which we include because often the memory function diverges at
the origin and $f_{0}\sim 1/\Delta t$, so this term is effectively of order $%
\Delta t.$ The additional density matrices are {\em virtual} in the sense
that they represent states that are never observed in the system behavior
and also in the related sense that they follow the paths of virtual photons.
This equation has the form of Eq.(\ref{tool}) and therefore the right hand
side can be put in the form of a superoperator acting on a density matrix.
In this way it is possible to propagate the non-Markovian evolution forward
in time.

Any expectation values are determined from $\rho _{n}^{\emptyset }$ alone; 
\begin{equation}
\langle a_{n}\rangle ={\rm Tr}\{a_{n}\rho _{n}^{\emptyset }\}.
\label{expect}
\end{equation}
For this reason it is convenient to introduce the projection operator $P$ $\ 
$defined by $\rho _{n}^{\emptyset }=P\underline{\rho }_{n},$ where $\rho
_{n}^{\emptyset }$ is projected out of the ``vector'' of matrices $%
\underline{\rho }_{n}$ which includes the reduced density matrix and all of
the virtual density matrices. In this notation we can write the expectation
value Eq.(\ref{expect}) as 
\begin{equation}
\langle a_{n}\rangle ={\rm Tr}\{a_{n}P\underline{\rho }_{n}\}
\end{equation}
and the evolution equation (\ref{final}) as 
\begin{equation}
\underline{\rho }_{n}=D_{n}\underline{\rho }_{n-1},
\end{equation}
where $D_{n}=D(\sigma _{n},\sigma _{n}^{\dagger },\sigma _{n}^{\prime },{%
\sigma _{n}^{\prime }}^{\dagger }).$ Often one is not only interested in
expectation values at a single time but also in temporal correlations. For
example, the correlation function (with $N_{2}>N_{1}$)

\begin{equation}
\langle \sigma _{N_{2}}^{\dagger }\sigma _{N_{1}}\rangle ={\rm Tr}\left\{
\sigma _{N_{2}}^{\dagger }\varrho _{N_{2},N_{1}}\right\}
\end{equation}
where, assuming that the initial state of the field is the vacuum state, we
have written

\begin{equation}
\varrho _{N_{2},N_{1}}=\sigma _{N_{1}}e^{{\cal L}(N_{2}\Delta t)}\rho .
\end{equation}
In general, techniques for calculating the correlation function such as the
quantum regression theorem (see, for example, \cite{carmichael}) will not
work in the non-Markovian case as there is no factorization of the evolution
matrix $e^{{\cal L}(N_{2}\Delta t)}$ at time $N_{1}\Delta t$. However, it is
not difficult to see that this type of correlation function is easily
calculated numerically within the above formalism. One simply propagates the
set of density matrices $\underline{\rho }_{n}$ forward from $n=0$ as usual;
then at $n=N_{1}$ multiply by $\sigma _{N_{1}}$; taking this as the new
initial state one can propagate up to time $N_{2}$ giving $\varrho
_{N_{2},N_{1}}$ to the accuracy of the algorithm. In symbolic notation this
procedure is written as

\begin{equation}
\varrho _{N_{2},N_{1}}\approx P\overleftarrow{\prod }
_{k=N_{1}+1}^{N_{2}}D_{k}\sigma _{N_{1}}\overleftarrow{\prod }
_{j=1}^{N_{1}}D_{j}\underline{\rho }_{0},
\end{equation}
where $\overleftarrow{\prod }$ denotes a time-ordered product.

For numerical simulations it is often more useful to work in the
Schr\"{o}dinger picture. Equation (\ref{final}) can easily be put in the
Schr\"{o}dinger picture \ because the operators are all acting at the same
time. We simply multiply by $U_{0}(n\Delta t){U_{0}^{\prime }}^{\dagger
}(n\Delta t)$ which transforms the density matrices at time $n$ to the
Schr\"{o}dinger picture and then write the interaction picture operators on
the right hand side in terms of the Schr\"{o}dinger picture operators. The
resulting equation has the same form as Eq.(\ref{final}) (with the density
matrices in the Schr\"{o}dinger picture) except the right hand side is
multiplied by $U_{0}(\Delta t){U_{0}^{\prime }}^{\dagger }(\Delta t)$; 
\begin{equation}
\underline{\rho }_{n}^{S}=D(\sigma ,\sigma ^{\dagger },\sigma ^{\prime },{%
\sigma ^{\prime }}^{\dagger })U_{0}(\Delta t){U_{0}^{\prime }}^{\dagger
}(\Delta t)\underline{\rho }_{n-1}^{S}.  \label{schrodingerpictureeqn}
\end{equation}
The advantage of the Schr\"{o}dinger picture is that the superoperator $D$
does not change with time and so only needs to be constructed once at the
start of the simulation.

It is worth noting that the above superoperator representation does not need
to be abandoned in numerical simulations. We can, for example, flatten the
density matrix into a column vector so that both the primed and unprimed
operators (which consequently become $4\times 4$ matrices rather than $%
2\times 2$ matrices) act only on the left-hand side of this flattened
density matrix \cite{tan99}. The right-hand side of Eq.(\ref
{schrodingerpictureeqn}) then represents the $4(3^{M})\times 4(3^{M})$
sparse matrix $D(\sigma ,\sigma ^{\prime },\sigma ^{\dagger },{\sigma
^{\prime }}^{\dagger })$ $U_{0}(\Delta t){U_{0}^{\prime }}^{\dagger }(\Delta
t)$ multiplying the $(3^{M})4\times 1$ column vector $\underline{\rho }
_{n}^{S}$. Note that the initial vector $\underline{\rho }_{0}$ has all
elements zero except for $\rho _{0}^{\emptyset }=\rho $. Also note that we
have written $M$ not $M-1$, as some virtual density matrices are initialized
at each time step.

\bigskip

\section{Results}

\label{sec:results}

Before we present the results of numerical simulations using the above
algorithm, it is important to make some general comments about the
limitations of the algorithm. As mentioned above, to propagate the
non-Markovian evolution requires a vector of size $(3^{M})4\times 1$ to be
updated at each time step, thus computer memory considerations put a severe
limitation on the number of time steps per memory time, $M$, that the
algorithm can deal with. In our case, we created the time independent
evolution matrix at the beginning and were in fact limited by the number of
non-zero entries of this matrix that could be stored in computer memory. On
a computer with 128MB of RAM we were limited to a maximum of $M=11$. This
obviously poses quite a problem when one wishes to consider quite
non-Markovian situations but also require high accuracy. However, for test
cases where the system was not strongly non-Markovian, the results of our
simulations show good agreement with calculations using standard methods of
analysis.

In the rest of this article we consider the case of on-resonance driving ($%
\omega _{0}=\omega _{L}$). In a frame rotating at the atomic frequency the
Hamiltonian for the atom reduces to 
\[
H_{0}=\frac{\hbar \Omega }{2}\;\left( \sigma ^{\dagger }+\sigma \right) . 
\]
This Hamiltonian gives rise to the following behavior 
\begin{eqnarray*}
\sigma (t) &=&e^{\frac{i}{\hbar }H_{0}t}\sigma e^{-\frac{i}{\hbar }H_{0}t} \\
&=&\frac{1}{2}\left( \sigma _{x}+|+\rangle \langle -|e^{i\Omega t}-|-\rangle
\langle +|e^{-i\Omega t}\right)
\end{eqnarray*}
where $\sigma _{x}=\sigma ^{\dagger }+\sigma $ and the states $|\pm \rangle $
are defined by $\sigma _{x}|\pm \rangle =\pm |\pm \rangle .$ When the atom
is coupled to a propagating field these three terms give rise to three
distinct peaks in the spectrum at $\omega -\omega _{0}=0,$ $\pm \Omega $.
This is referred to as the resonance fluorescence triplet \cite{mollow,resfl}%
.

With a structured electromagnetic field, these emission peaks may be
altered. In the following sections we use our algorithm to consider the case
when the atom is in a cavity tuned to one of the side peaks and the case
when one of the side-peaks is at a frequency where no photons propagate in a
bandgap material.

\subsection{Atom in a Cavity}

\label{sec:atomincavity}

In order to demonstrate the validity of our algorithm we consider a simple
system that can also be described by a Born-Markov master equation (and so
can be solved by traditional methods) and compare the results of the two
methods. The system that we consider is an atom coupled to a single mode of
a cavity which is in turn coupled to a continuum of propagating modes of the
electromagnetic field via a partially transmissive mirror. The standard
method of finding the evolution of this system is to treat the atom and the
cavity as a system and the external field as a reservoir which can be traced
over. This produces a Born-Markov master equation for the atom and cavity
modes. This can then be solved by standard methods \cite{carmichael}. One
can also make a different choice of system and reservoir; consider the
evolution of the atom alone while treating both the cavity and the external
field as the reservoir; giving rise to a non-Markovian problem.

The usual description of this system is in terms of the Hamiltonian 
\begin{equation}
H_{1}=H_{0}+\widetilde{H}+i\hbar \sqrt{\frac{\gamma }{2\pi }}\left[
a^{\dagger }\sigma -\sigma ^{\dagger }a\right] ,  \label{proofofequivHone}
\end{equation}
where 
\begin{equation}
\widetilde{H}=\hbar \nu a^{\dagger }a+\hbar \int_{-\infty }^{\infty }d\omega
\quad \left\{ \omega b^{\dagger }(\omega )b(\omega )+i\frac{\kappa }{\sqrt{%
2\pi }}\left[ b^{\dagger }(\omega )a-b(\omega )a^{\dagger }\right] \right\} ,
\label{proofofequivHprime}
\end{equation}
and where $a$ is the annihilation operator of the cavity mode, $b(\omega )$
is the annihilation operator for the external propagating electromagnetic
field mode of energy $\hbar \omega $, $\sigma $ is the lowering operator for
the atom, and $H_{0}$ acts only on the atomic operators.

Choosing the atom as the system and eliminating the cavity mode it is
possible to describe this system by the alternative Hamiltonian (see
Appendix. \ref{app:atomincavity} ) 
\begin{equation}
H_{2}=H_{0}+\hbar \int_{-\infty }^{\infty }d\omega \quad \left\{ \omega
c^{\dagger }(\omega )c(\omega )+i[\lambda ^{\ast }(\omega )c^{\dagger
}(\omega )\sigma -\lambda (\omega )c(\omega )\sigma ^{\dagger }]\right\} ,
\label{alternativeH}
\end{equation}
where $c(\omega )$ is a new bose annihilation operator and the, now
frequency dependent, coupling constant is 
\begin{equation}
\lambda (\omega )=\sqrt{\frac{\gamma }{2\pi }}\frac{\kappa }{i(\omega -\nu )-%
\frac{\kappa ^{2}}{2}}.  \label{cavitycoupling}
\end{equation}
The memory function in this case is therefore given by 
\begin{eqnarray}
f_{{\rm m}}(t-s) &=&\int_{-\infty }^{\infty }d\omega \int_{-\infty }^{\infty
}d\omega ^{\prime }\quad \lambda ^{\ast }(\omega ^{\prime })\lambda (\omega
)[c(\omega ),c^{\dagger }(\omega ^{\prime })]e^{-i\omega t+\omega ^{\prime
}s} \\
&=&\frac{\gamma }{2\pi }\int_{-\infty }^{\infty }d\omega \frac{\kappa ^{2}}{%
(\omega -\nu )^{2}+\left( \frac{\kappa ^{2}}{2}\right) ^{2}}e^{-i\omega
(t-s)} \\
&=&\gamma \exp \left\{ -i\nu (t-s)-\frac{\kappa ^{2}}{2}|t-s|\right\} .
\end{eqnarray}
Transforming to a frame rotating at the atomic frequency appends a factor $%
e^{i\omega _{0}(t-s)}$ to this memory function. This model differs from that
of the bandgap only by the functional form of the memory function.

We compare our results for the system described by $H_{2}$ using the
non-Markovian algorithm introduced in Sec. \ref{sec:method} with the
solution of the Born-Markov master equation for the atom plus cavity system
as described by $H_{1}$. In Fig.\ref{fig:cavityavg} we plot the time
evolution of the probability for the atom to be in the excited state. The
internal spectrum, $S(\omega )$, defined as the Fourier transform of the
steady-state correlation function $\langle \sigma ^{\dagger }(\tau )\sigma
\rangle _{{\rm ss}}-\langle \sigma ^{\dagger }\rangle _{{\rm ss}}\langle
\sigma \rangle _{{\rm ss}}$ , is plotted in Fig.\ref{fig:cavityspec} showing
a significant asymmetry in the height of the side peaks due to the tuning of
the cavity to only one side band of the resonance fluorescence triplet. This
is a completely non-Markovian effect as it arrises directly from the
structured nature of the field, i.e., the fact that some frequencies of the
field are easier to emit into than others. Both the probability and the
spectrum show good agreement with the results of the traditional method (we
used the numerical routines in the quantum optics toolbox \cite{tan99}).

\bigskip

\subsection{A Bandgap Material}

In this section we apply our algorithm to resonance fluorescence in a
bandgap material; a problem that cannot be solved by any traditional
methods. Following the discussion of Vats and John \cite{vats} we consider
an {\em anisotropic} bandgap model under the effective mass approximation.
This approximation is appropriate to fabricated band gap materials. In this
model the band edge is associated with a specific point in $k$-space, ${\bf %
k=k}_{0}$, and the dispersion relation close to the upper edge of the gap
has the form

\begin{equation}
\omega _{{\bf k}}=\omega _{g}+\frac{\hbar \left( {\bf k-k}_{0}\right) ^{2}}{%
2\mu}.
\end{equation}
where $\mu$ is called the effective mass in analogy with the dispersion
relations of massive particles.

This dispersion relation gives rise to a memory function with the asymptotic
behavior $f_{{\rm m}}(\tau )\sim \tau ^{-3/2}$ so that, in contrast to an 
{\em isotropic} bandgap (where $f_{{\rm m}}(\tau )\sim \tau ^{-1/2})$ we are
able to define a finite memory time.

The memory function given by $f_{{\rm m}}(\tau )\propto \tau ^{-3/2}$
suffers from a singularity at $\tau =0.$ This is due to the absence of a
high-frequency cutoff, which must be included to model a physical system. We
include this high frequency cutoff \ by considering the slightly modified
memory function,

\begin{equation}
f_{{\rm m}}(\tau )=\beta \lambda ^{3/2}\frac{e^{i\left[ \delta \tau +\pi /4 %
\right] }}{\left( 1+\lambda \tau \right) ^{3/2}},  \label{bandgapmf}
\end{equation}
where (shifting to a frame rotating at the atomic frequency) $\delta =\omega
_{0}-\omega _{g}$ is the detuning between the band edge and the atomic
resonance frequency and $\beta $ is a measure of the coupling strength
between the atom and the electromagnetic field. The parameter $\lambda$ is
infinite when there is no upper frequency limit, so for large $\lambda $
this memory function is effectively the same as that derived by Vats and
John. It is useful to introduce the quantity

\begin{eqnarray*}
\gamma &=&2\Re \int_{0}^{\infty }d\tau \quad \left. f_{{\rm m}}(\tau
)\right| _{\delta =0} \\
&=&2^{3/2}\beta \sqrt{\lambda }
\end{eqnarray*}
which corresponds to the Born-Markov damping rate for this memory function.
We use this quantity as a scale against which we can measure the various
rates.

For large $\lambda ,$ the factor $\left( 1+\lambda \tau \right) ^{-3/2}$ has
a sharply rising behavior as $\tau \rightarrow 0$ on a time scale much
faster than the system variables. This is a common feature of physical
memory functions and requires special treatment in numerical calculations.
We generalize the trapezoidal rule used for the atom in a cavity case by
calculating weights that depend on the first two moments of $\left(
1+\lambda \tau \right) ^{-3/2}$ and derive an extended two-point integration
scheme on a uniform mesh as described in Ref.\cite{numerical}. \ Explicitly,
we used the weights

\[
W_{n}=\left\{ 
\begin{array}{cc}
w_{0}^{0}-w_{1}^{0}, & n=0 \\ 
-(M-2)w_{0}^{M-2}+w_{1}^{M-2}, & n=M-1 \\ 
(n+1)w_{0}^{n}-w_{1}^{n}-(n-1)w_{0}^{n-1}+w_{1}^{n-1},\quad & {\rm otherwise}
\end{array}
\right. 
\]
where 
\[
w_{j}^{n}=\frac{1}{\Delta t^{j}}\int_{n\Delta t}^{(n+1)\Delta t}d\tau \quad
\tau ^{j}\left( 1+\lambda \tau \right) ^{-3/2}. 
\]
Since 
\[
\sum_{n=0}^{n=M-1}W_{n}=\int_{0}^{(M-1)\Delta t}d\tau \quad \left( 1+\lambda
\tau \right) ^{-3/2}, 
\]
this integration scheme treats the semi-singular part of the memory function
exactly and allows us to treat a range of non-Markovian situations,
including the Markovian limit, within the same algorithm. As mentioned
previously we do not consider schemes of higher order than the two-point
scheme as the numerical difficulty of evolving the system a single timestep
increases dramatically due to the combinatorics introduced by the
time-ordering.

Let us first consider the undriven case where an initially excited atom
simply undergoes decay inside a bandgap material. Since this case can be
treated by an alternative method it serves as another test of the algorithm.
In the case of simple decay there is at most only one photon in the
electomagnetic field so we can write the state of the atom and field as 
\[
|\Psi (t)\rangle =a(t)|1\rangle |\{0\}{\bf \rangle +}\sum_{k}c_{k}(t)b_{k}^{%
\dagger }|0\rangle |\{0\}{\bf \rangle ,} 
\]
where $|1\rangle $ and $|0\rangle $ are the excited and ground states of the
atom and $|\{0\}{\bf \rangle }$ is the field vacuum. It is not difficult to
establish that 
\[
\frac{da(t)}{dt}=-\int_{0}^{t}ds\quad f_{{\rm m}}(t-s)a(s), 
\]
with $f_{{\rm m}}(t-s)$ given by Eq.(\ref{bandgapmf}). This
integro-differential equation can be numerically simulated by, for example,
the routine (\ref{routine}). In actual fact we used the weighting scheme
described above to approximate the integral. The probability of the atom
being in the excited state is given by $P(t)=|a(t)|^{2}.$ In Fig.\thinspace 
\ref{fig:decay} we have compared the results of our algorithm to a direct
numerical integration of the above integro-differential equation; the two
methods show good agreement. In contrast to the isotropic band gap case,
previously treated analytically in Ref.\thinspace \cite{john94}, the decay
is always exponential. This can be seen more clearly in the corresponding
plot of the logarithm of $P(t)$, Fig.\thinspace \ref{fig:decaylog}. Note
that in this plot the discrepancy between the two methods is apparent. Our
algorithm gets more accurate as $\lambda $ is increased as this allows a
smaller stepsize to be taken. These plots were made with the parameter value 
$\lambda =3\times 10^{2}\gamma $ and the variation of the damping rate with
detuning from the bandgap is evident. However, when this parameter is
increased the situation becomes more and more Markovian, despite the long
time tail of the memory function. For $\lambda =10^{5}\gamma $ there is no
noticeable difference from the Markovian case; the decay rate is exactly
that predicted by the Born-Markov result, and the detuning becomes
irrelevant. This is shown in Fig.\thinspace \ref{fig:decayMarkov}. Note that
the $\delta <0$ case requires the driving laser to be propagating in a
direction that is perpendicular to the direction of $\ {\bf k}_{0}$ for the
anisotropic bandgap.

For non-zero driving there are no traditional methods available and the
results can only be determined by a quantum non-Markovian treatment. In
Fig.\thinspace \ref{fig:bandgapavg} we have plotted the probability
amplitude for the atom to be in its excited state. The driving induces
oscillations in this probability and the atom tends to a steady state close
to equal probability of being in the excited state and in the ground state
due to the reasonably large driving. This behavior is qualitatively
equivalent to the Markovian case. The steady-state correlation function, $%
C(\tau )=\langle \sigma ^{\dagger }(\tau )\sigma \rangle _{{\rm ss}}-\langle
\sigma ^{\dagger }\rangle _{{\rm ss}}\langle \sigma \rangle _{{\rm ss}},$ is
plotted in Fig.\thinspace \ref{fig:bandgapcorr}. Taking the Fourier
transform of the steady-state correlation function, $C(\tau ),$ gives the so
called internal spectrum of the atom, $S(\omega ).$ As in the atom in a
cavity case this quality allows us to clearly see the effects of the
non-trivial structure in the field on the behavior of the atom. In
Fig.\thinspace \ref{fig:bandgapspec} we have plotted the internal spectrum
with $\lambda =3\times 10^{2}\gamma $ for various detunings from the band
edge. A definite asymmetry arises in the magnitude of the side peaks
directly due to the fact that emission at frequencies inside the gap is
prohibited. Note that as $\lambda $ is increased this effect becomes less
and less until the spectrum becomes equivalent to the Markovian case. In Fig.%
\ref{fig:bandgapspec2} we have plotted the internal spectrum with the same
detunings but with $\lambda =10^{5}\gamma ;$ there is no evidence of
asymmetry and the values of detuning become irrelevant just as in the case
of the undriven atom.

\section{Conclusions}

\label{sec:conclusions}

A number of different methods of dealing with non-Markovian systems have
appeared in the literature recently. These methods can be broadly separated
into two main categories: methods that involve extending the system size by
the addition of real or imaginary modes \cite{imamoglu,garraway,garraway2}
such that the new extended system evolves via a Born-Markov master equation,
and methods that attempt to evolve the non-Markovian evolution directly \cite
{jack00,diosi98,struntz99,breuer99,jack99}. Note that the approach in Ref. 
\cite{breuer99} takes advantage of the special case when the free evolution
of the system has an analytical solution to derive a Markov equation for the
system. Conceptually, the present work falls into the last category,
however, in order to propagate the non-Markovian evolution we have
effectively extended the system size by introducing virtual density matrices
that follow the paths of the photons in the field before they are either
irretrievably emitted or are reabsorbed by the atom. The two approaches have
very different origins and the way in which the extended system is
constructed is fundamentally different. The present method seems to be
slightly more general as no restrictions are made on the form of the memory
function, however, time discretization plays a much more fundamental role in
our algorithm so it would be very interesting to see on what level these two
approaches are related.

The strength of the present algorithm lies in its ability to accurately
model weakly non-Markovian systems. An appropriate application would be to
determining the validity of the Born-Markov approximation in borderline
cases. The biggest constraint on the algorithm is the power law increase in
the number of virtual density matrices that need to be stored to propagate
the non-Markovian evolution. This is a fundamental constraint and arises
from the inherent difficulty in numerically simulating quantum mechanical
systems in general. Note that the algorithm presented here was derived
without using the algebraic relations of the atomic operators. It is
therefore applicable to other system-bath couplings that are linear in the
bath operators.

With our algorithm we have determined the evolution of a driven atom
radiating into a bandgap material with an anisotropic bandgap; a situation
that cannot be treated by any traditional methods. The asymmetry in the
internal spectrum of the atom clearly demonstrates the effect of the
non-trivial structure of the field on the behavior of the atom. This
behavior cannot occur under the Born-Markov approximation. In addition, we
have demonstrated the transition to the Born-Markov limit, which occurs
despite the presence of a long time tail on the memory function.

In this article we have presented a density matrix version of the algorithm.
In this case the system size was so small that a density matrix treatment
was appropriate, however, a wave function version can also be derived in a
similar fashion. The wavefunction version is applicable to simulating
non-Markovian quantum trajectories that also solve for the reduced density
matrix \cite{jack00,diosi98,struntz99,jack99}. In the non-Markovian case
increasing system size is obviously an even more difficult constraint to
numerical commutation than in the Markovian case so wave function methods
will be important.

\acknowledgements
The authors would like to thank M. Collett for useful discussions and M.J.
would like to acknowledge the support of the Japan Society for the Promotion
of Science. J.H. was supported by the University of Auckland Research
Committee and the Marsden Fund of the Royal Society of New Zealand.

\bigskip

\appendix

\section{Non-Markovian model for an atom in a cavity}

\label{app:atomincavity}

This appendix shows how the system in section \ref{sec:atomincavity} can be
written in terms of a non-Markovian model.

Equation (\ref{proofofequivHprime}) is valid in the high-frequency limit 
\cite{collett85}: it assumes the rotating wave approximation and that the
fundamental frequency $\nu $ is much greater than the spread of frequencies
about the fundamental frequency such that we can replace the lower bound of
the integral over the bath modes by $-\infty$. Note that as we are
considering a one dimensional field in the continuum limit it is convenient
to work in terms of the frequencies rather than the mode labels.

The proof that the two Hamiltonians (\ref{proofofequivHone}) and (\ref
{alternativeH}) are equivalent consists in demonstrating that they are
related by a unitary transformation. Let the initial time of interest be $%
t=0 $. \ For the moment let us restrict our consideration to the cavity
mode-field system described by the Hamiltonian $\widetilde{H}$ [see Eq.(\ref
{proofofequivHprime})]. Consider the following scenario: let the coupling
between the cavity and the field be time dependent and let it increase
slowly such that at a long time in the past $(t<0)$ the coupling is turned
off and at the initial time $(t=0)$ it is equal to $\kappa$, 
\begin{equation}
\kappa (t)=\left\{ 
\begin{array}{cc}
\kappa e^{\epsilon t}, & t<0 \\ 
\kappa , & t\geq 0
\end{array}
\right. ,
\end{equation}
where $\ \epsilon $ is a small positive number. The adiabatic limit of an
infinitely slow switching on of the coupling $\ $is given by the limit where 
$e^{\epsilon t}\rightarrow 0$ as $t\rightarrow -\infty .$ and $\epsilon
\rightarrow 0$ The Heisenberg equations of motion for the cavity and field
modes for $t<0$ are given by

\begin{eqnarray}
\frac{d\widetilde{a}(t)}{dt} &=&-\frac{i}{\hbar }[\widetilde{a}(t),%
\widetilde{H}(t)]=(-i\nu -\epsilon )\widetilde{a}(t)-\frac{\kappa }{\sqrt{%
2\pi }}\int_{-\infty }^{\infty }d\omega \quad b(\omega ,t)
\label{proofofequivprimeone} \\
\frac{db(\omega ,t)}{dt} &=&-\frac{i}{\hbar }[b(\omega ,t),\widetilde{H}%
(t)]=-i\omega b(\omega )+\frac{\kappa }{\sqrt{2\pi }}\widetilde{a}(t),
\label{proofofequivprime2}
\end{eqnarray}
where we have simplified the equations by making the replacement $\widetilde{%
a}(t)=a(t)e^{\epsilon t}.$ Note that if ${\rm \lim }_{t_{0}\rightarrow
-\infty }$ $a(t_{0})$ is finite this implies that

\begin{equation}
{\rm \lim_{t_{0}\rightarrow -\infty }}\widetilde{a}(t_{0})=0.
\label{needed1}
\end{equation}
Solving Eq.(\ref{proofofequivprime2}) formally for $b(\omega ,t)$ in terms
of $b(\omega ,t_{0}),$ where $t_{0}<t,$ and substituting this solution into
Eq. (\ref{proofofequivprimeone}) we derive

\begin{equation}
\frac{d\widetilde{a}(t)}{dt}=-\left( i\nu +\epsilon +\frac{\kappa ^{2}}{2}%
\right) \widetilde{a}(t)-\frac{\kappa }{\sqrt{2\pi }}\int_{-\infty }^{\infty
}d\omega \quad b(\omega ,t_{0})e^{-i\omega (t-t_{0})},
\end{equation}
where we have used the relation 
\begin{equation}
\delta (t-s)=\frac{1}{2\pi }\int_{-\infty }^{\infty }d\omega \quad
e^{-i\omega (t-s)}.
\end{equation}
Solving this formally for $\widetilde{a}(0)$ in terms of $\widetilde{a}%
(t_{0})$ we find

\begin{equation}
\widetilde{a}(0)=\widetilde{a}(t_{0})e^{(i\nu +\epsilon +\frac{\kappa ^{2}}{2%
})t_{0}}-\frac{\kappa }{\sqrt{2\pi }}\int_{-\infty }^{\infty }d\omega \quad 
\widetilde{b}(\omega ,t_{0})\frac{1-e^{[i(\nu -\omega )+\epsilon +\frac{%
\kappa ^{2}}{2}]t_{0}}}{\left[ i(\nu -\omega )+\epsilon +\frac{\kappa ^{2}}{2%
}\right] }.
\end{equation}
where $\widetilde{b}(\omega ,t_{0})=$ $b(\omega ,t_{0})e^{i\omega t_{0}}.$
Letting $t_{0}\rightarrow -\infty $ and $\epsilon \rightarrow 0$ (and
assuming that $\widetilde{b}(\omega ,t_{0})$ exists in this limit) this
solution becomes 
\begin{equation}
a(0)=\widetilde{a}(0)=\frac{1}{\sqrt{2\pi }}\int_{-\infty }^{\infty }d\omega
\quad \widetilde{b}(\omega ,-\infty )\frac{\kappa }{\left[ i(\omega -\nu )-%
\frac{\kappa ^{2}}{2}\right] }.  \label{needed2}
\end{equation}
As the system contains no bound states the Hamiltonian in the same limit is
given by 
\begin{equation}
\widetilde{H}(-\infty )=\hbar \int_{-\infty }^{\infty }d\omega \quad \omega 
\widetilde{b}^{\dagger }(\omega ,-\infty )\widetilde{b}(\omega ,-\infty ).
\label{needed3}
\end{equation}
Therefore, the unitary transformation given by 
\begin{equation}
\widetilde{U}=\lim_{t_{0}\rightarrow -\infty ,\epsilon \rightarrow 0}\left[ 
\overrightarrow{T}\exp \left\{ \frac{i}{\hbar }\int_{t_{0}}^{0}ds\quad 
\widetilde{H}(s)\right\} \right] .
\end{equation}
diagonalizes the Hamiltonian $\widetilde{H}(0)$ and gives $a(0)$ in terms of
the diagonalizing modes. From Eq.(\ref{needed1}), (\ref{needed2}) and (\ref
{needed3}) we see that this is just the transform required to write the
Hamiltonian Eq.(\ref{proofofequivHone}) as Eq.(\ref{alternativeH}) with $%
c(\omega )=\widetilde{b}(\omega ,-\infty ),$ thus demonstrating their
physical equivalence.

\bigskip

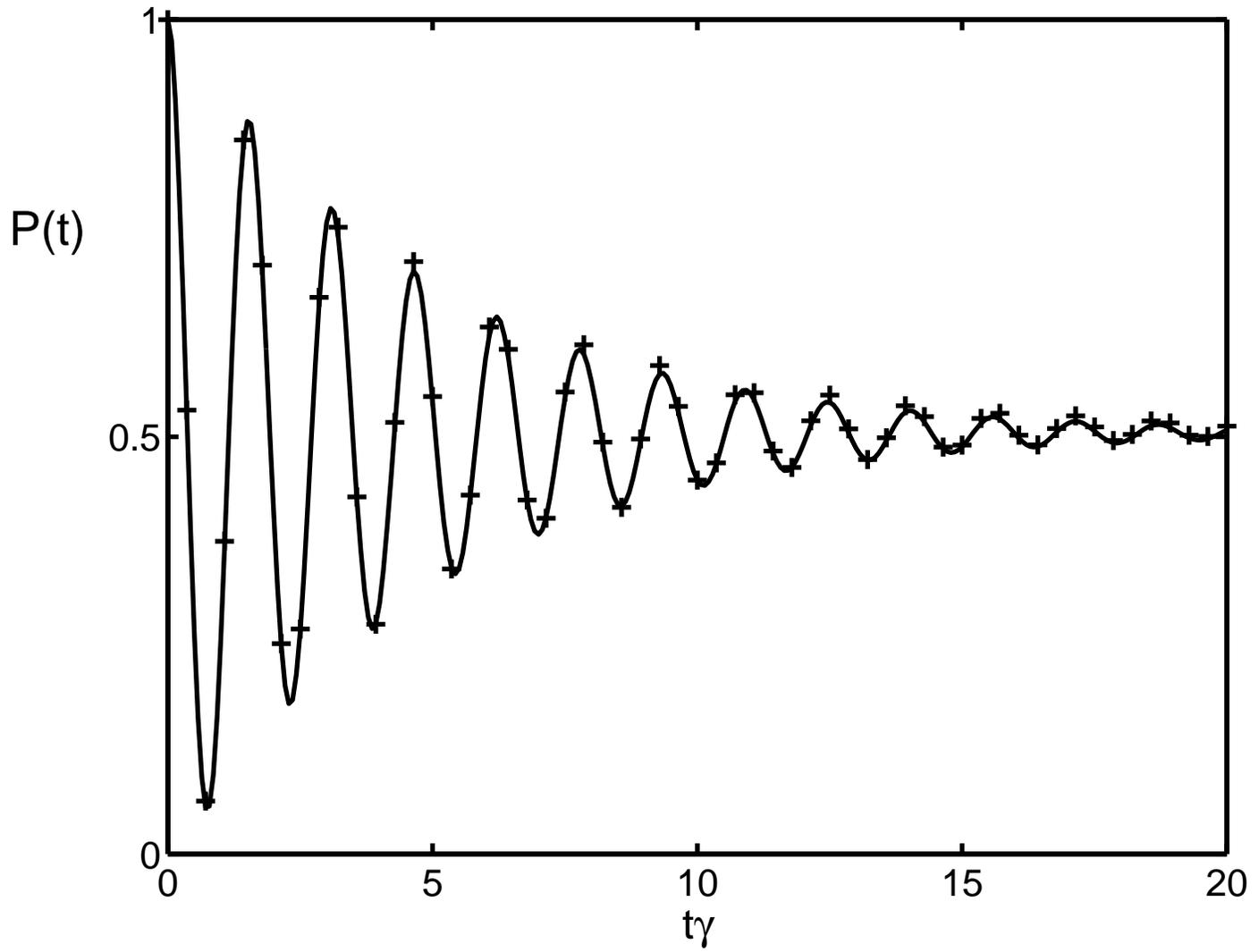
\begin{figure}[tbp]
\caption{Time evolution of $P(t)$ from an initial excitted state for the
parameters: $\Omega =4\protect\gamma$, $\protect\omega_{0}-\protect\nu
=\Omega $ and $\protect\kappa ^{2}=8\protect\gamma $. For the non-Markovian
algorithm we used a memory time of $M=11$ and a time step of \ $\Delta
t=1/14 $. The solid line is the result of the non-Markovian algorithm and
the crosses the result of a numerical solution of the master equation for
the extended system of atom plus cavity mode. }
\label{fig:cavityavg}
\end{figure}
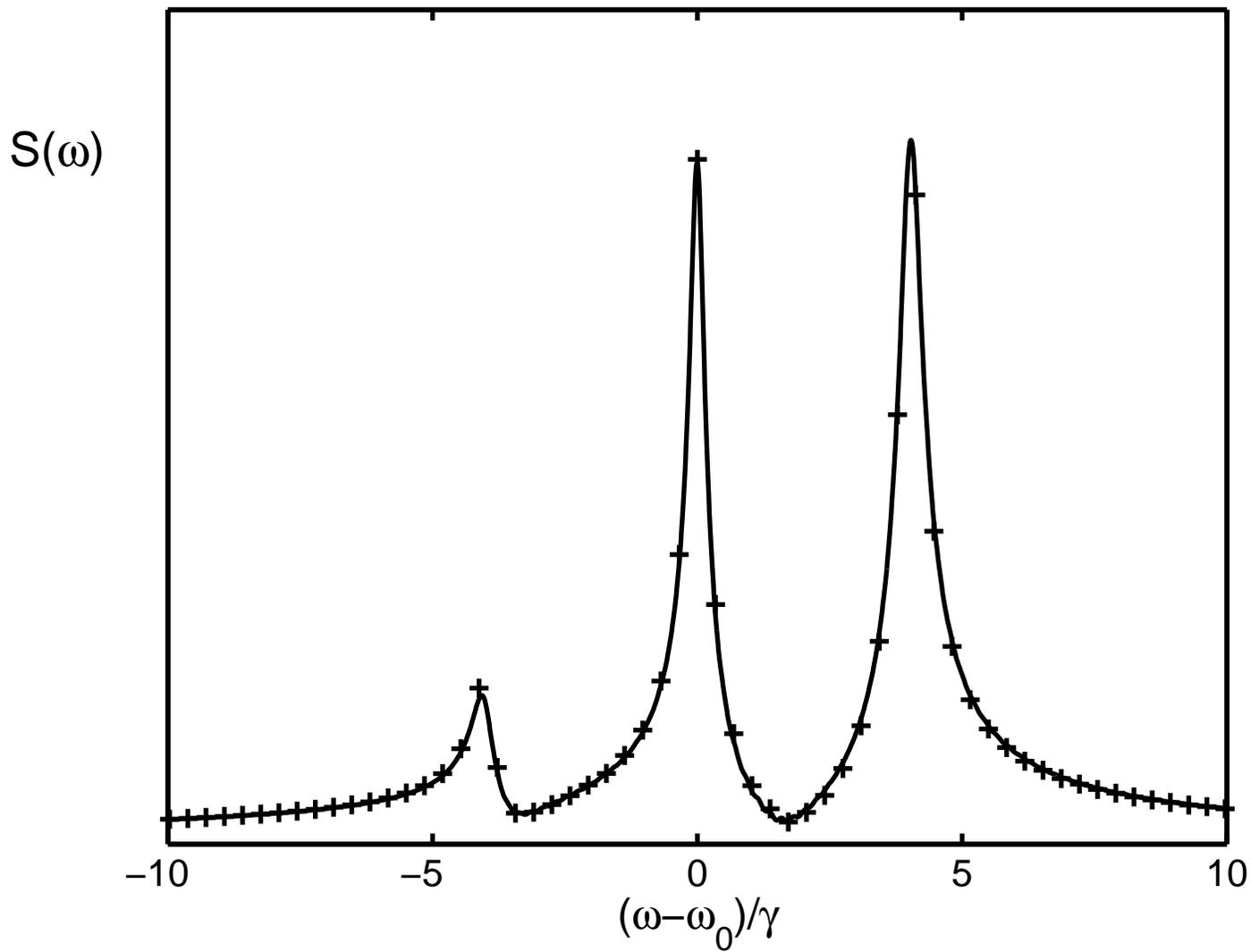
\begin{figure}[tbp]
\caption{Internal spectrum, $S(\protect\omega )$, for the same parameters as
Fig.1. Again the solid line is the result of the non-Markovian algorithm and
the crosses the results of a numerical solution of the master equation for
the extended system.}
\label{fig:cavityspec}
\end{figure}
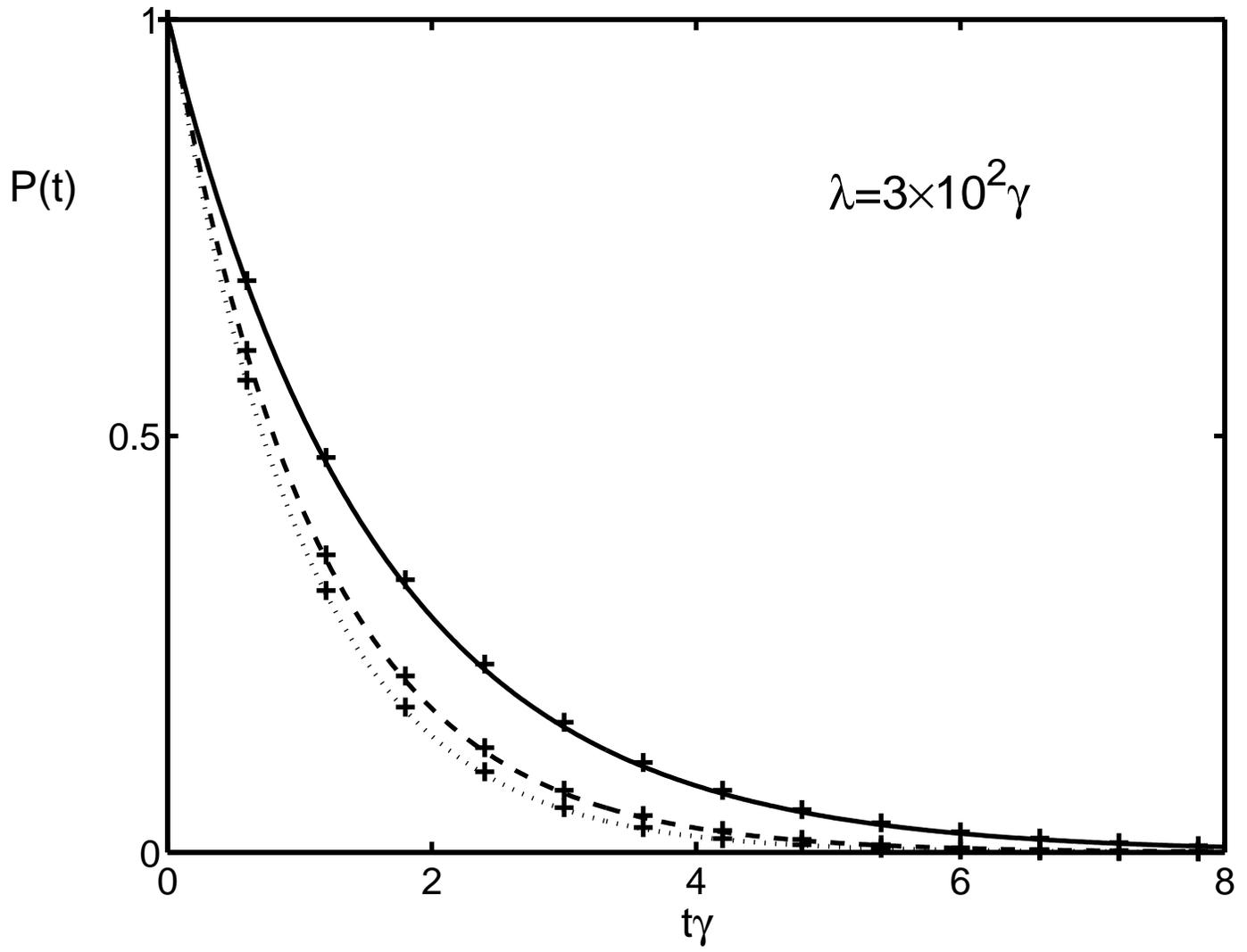
\begin{figure}[tbp]
\caption{Plot of $P(t)$ for an undriven atom initially in an excitted state
for different detunings from the band edge. Parameter values are $\protect%
\lambda=3\times 10^{2}\protect\gamma $, $M=11$ and $\Delta t=1/50$. The
various detunings in the plot are $\protect\delta =10\protect\gamma $ (solid
line), $\protect\delta=0$ (dashed line) and $\protect\delta =-10\protect%
\gamma $ (dotted line).}
\label{fig:decay}
\end{figure}
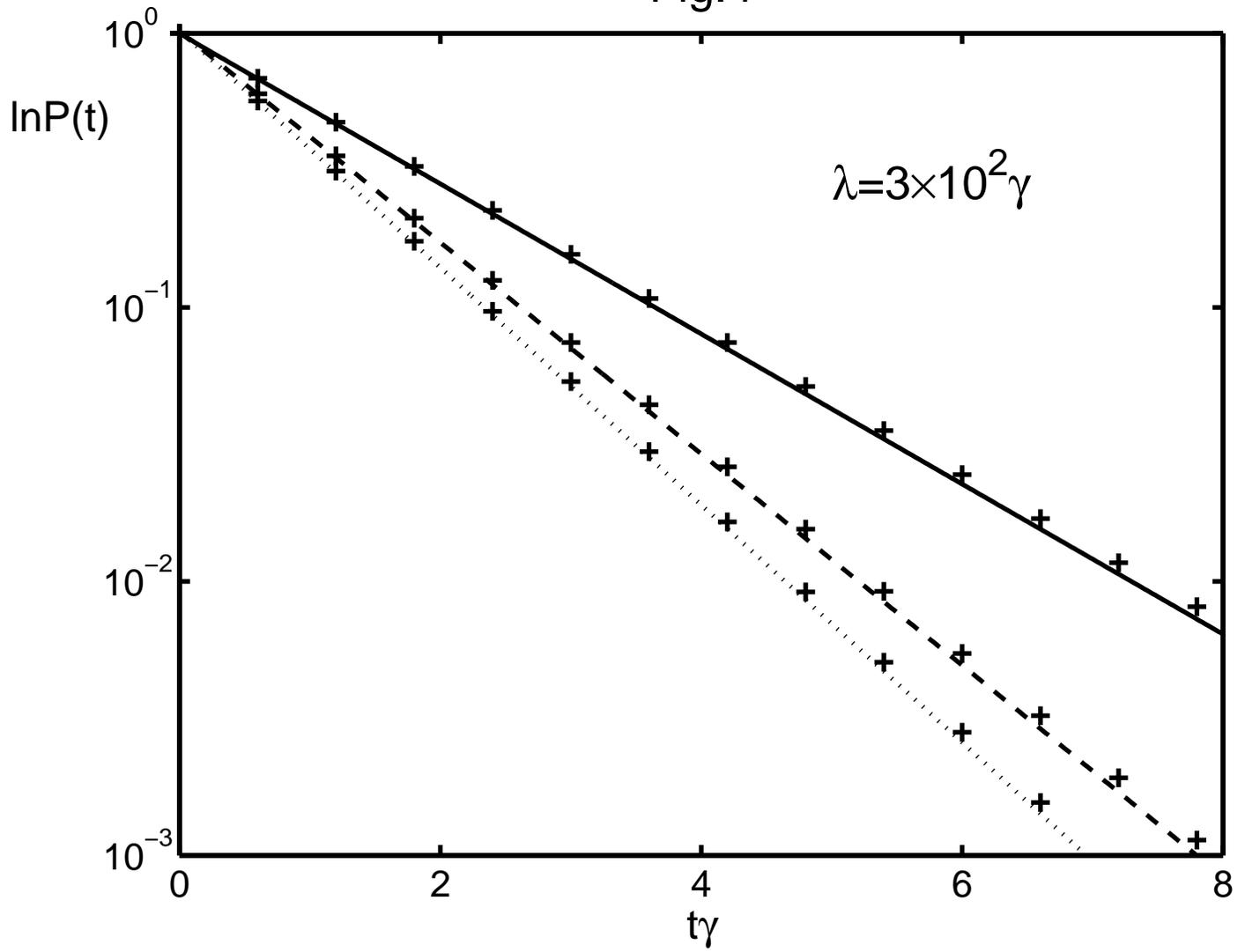
\begin{figure}[tbp]
\caption{Same as Fig.3 but with $P(t)$ on a logrithmic scale. The straight
lines indicate no deviation from exponential decay.}
\label{fig:decaylog}
\end{figure}
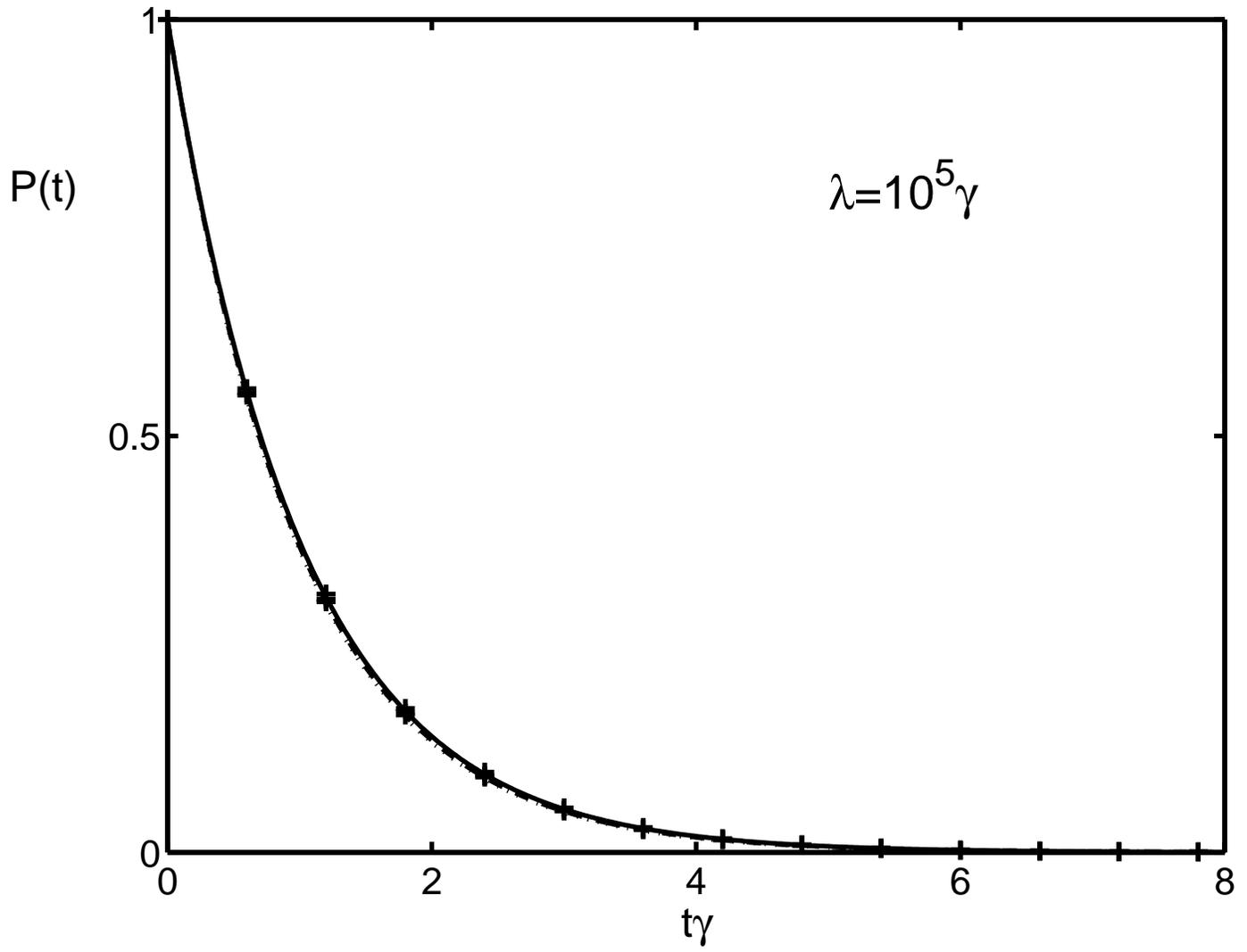
\begin{figure}[tbp]
\caption{Plot of $P(t)$ for $\protect\lambda =10^{5}\protect\gamma $ with
other parameters the same as Fig.3. The lines corresponding to different
detunings are indistinguishable. }
\label{fig:decayMarkov}
\end{figure}
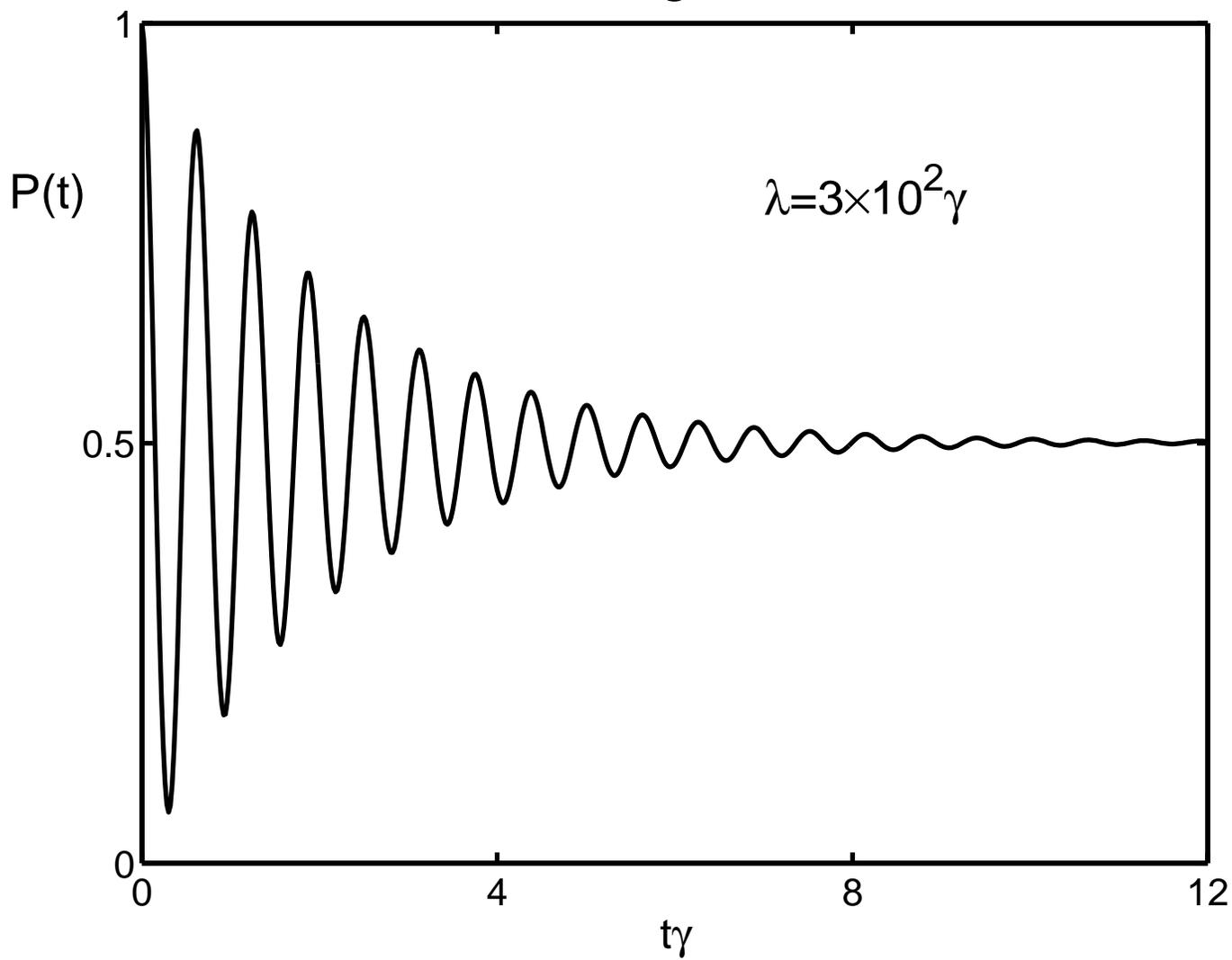
\begin{figure}[tbp]
\caption{Plot of $P(t)$ for a driven atom for the parameter values: $%
\Omega=10\protect\gamma$, $\protect\lambda =3\times 10^{2}\protect\gamma $, $%
M=11$, $\Delta t=1/50$ and $\protect\delta =10\protect\gamma $. }
\label{fig:bandgapavg}
\end{figure}
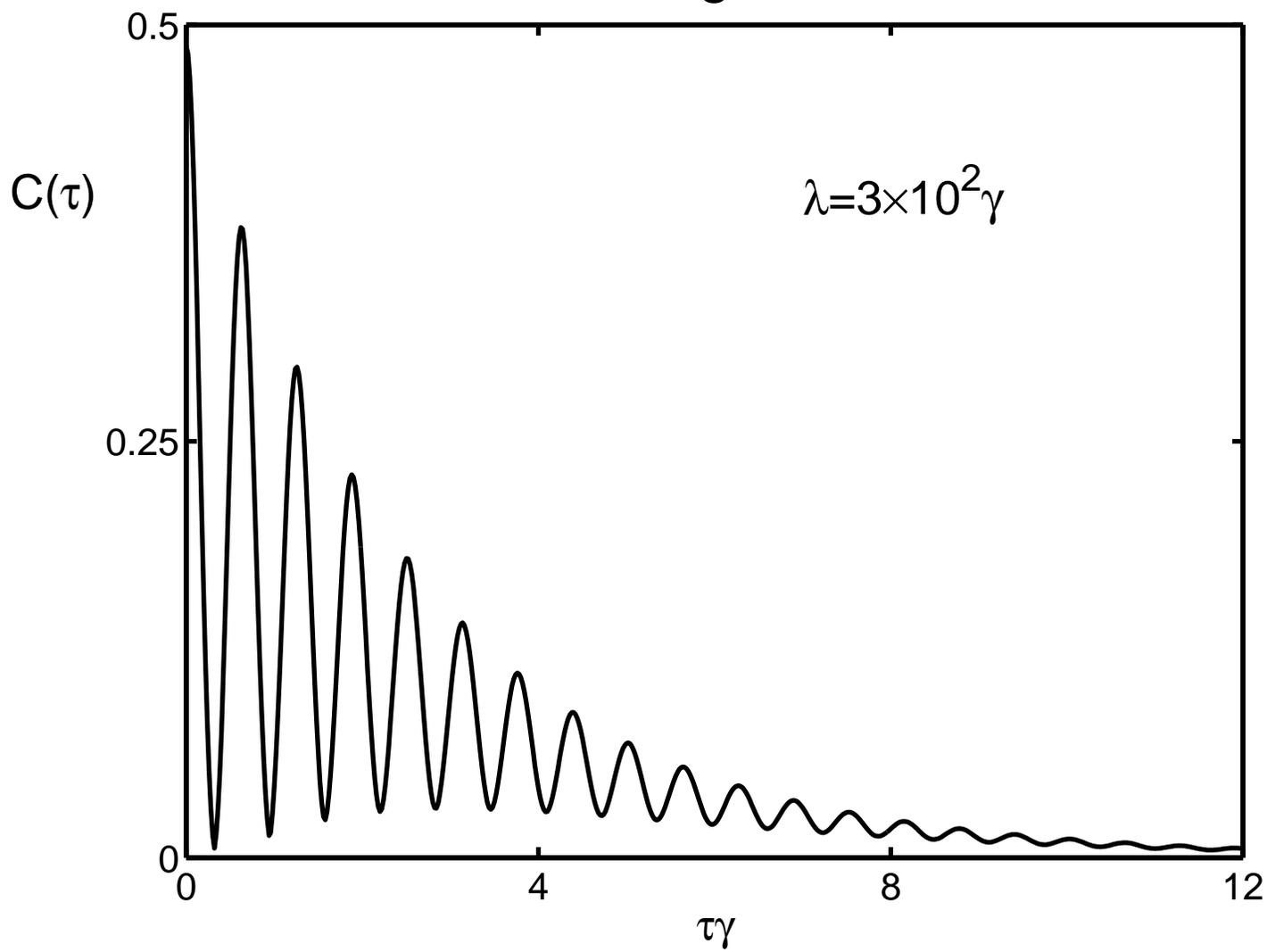
\begin{figure}[tbp]
\caption{Steady state correlation function, $C(\protect\tau)=\langle\protect%
\sigma^{\dagger}(\protect\tau)\protect\sigma\rangle_{{\rm ss}} -\langle%
\protect\sigma^{\dagger}\rangle_{{\rm ss}}\langle\protect\sigma\rangle_{{\rm %
ss}}$, of the atom with the same parameters as Fig.6.}
\label{fig:bandgapcorr}
\end{figure}
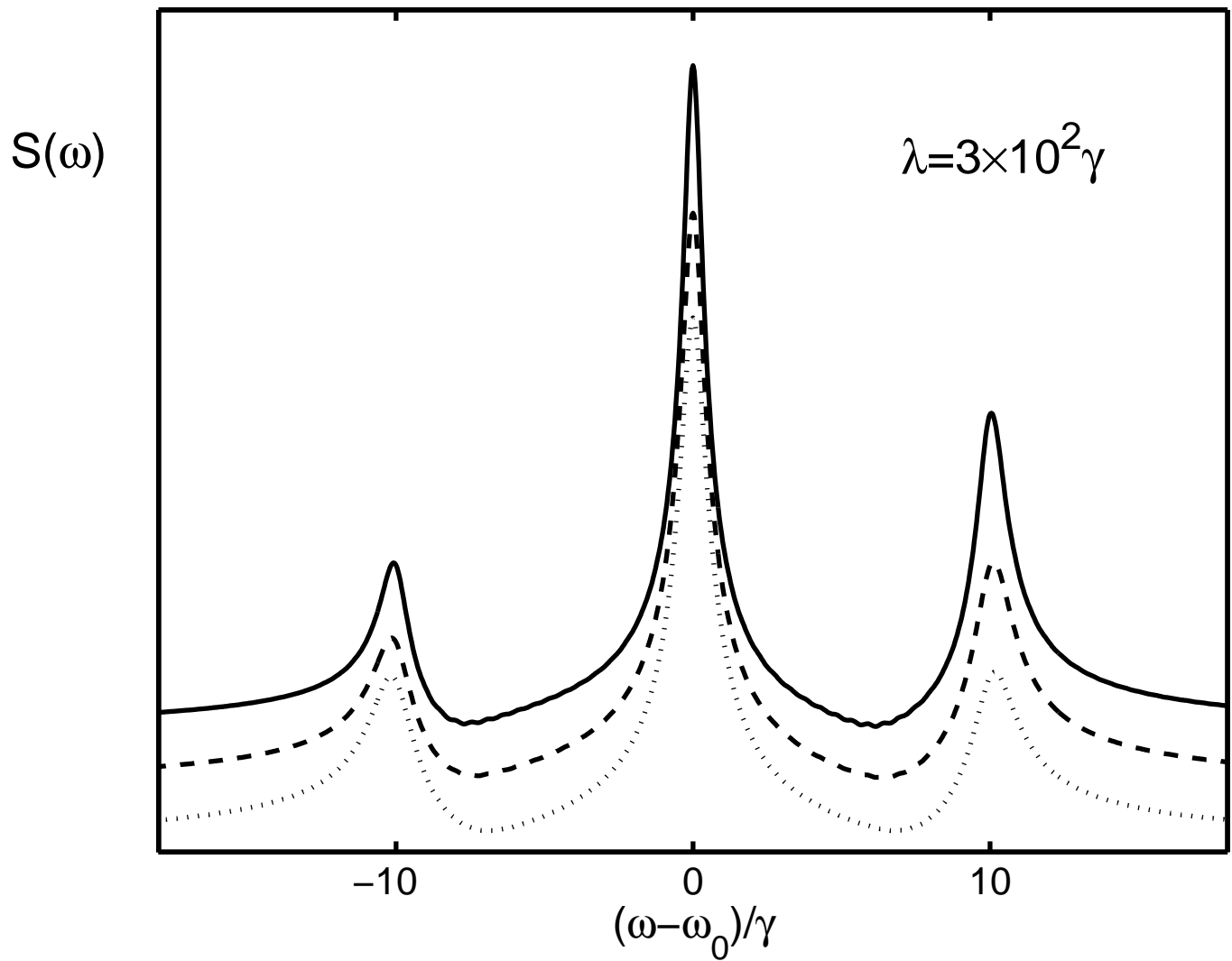
\begin{figure}[tbp]
\caption{Internal spectrum, $S(\protect\omega)$, for $\Omega=10\protect\gamma
$, $\protect\lambda =3\times10^{2}\protect\gamma$, $M=11$, and $\Delta t=1/50%
$. The various spectra at different detunings have been displaced for ease
of comparison. The detunings are $\protect\delta =10\protect\gamma $ (solid
line), $\protect\delta =0 $ (dashed line) and $\protect\delta =-10\protect%
\gamma $ (dotted line).}
\label{fig:bandgapspec}
\end{figure}
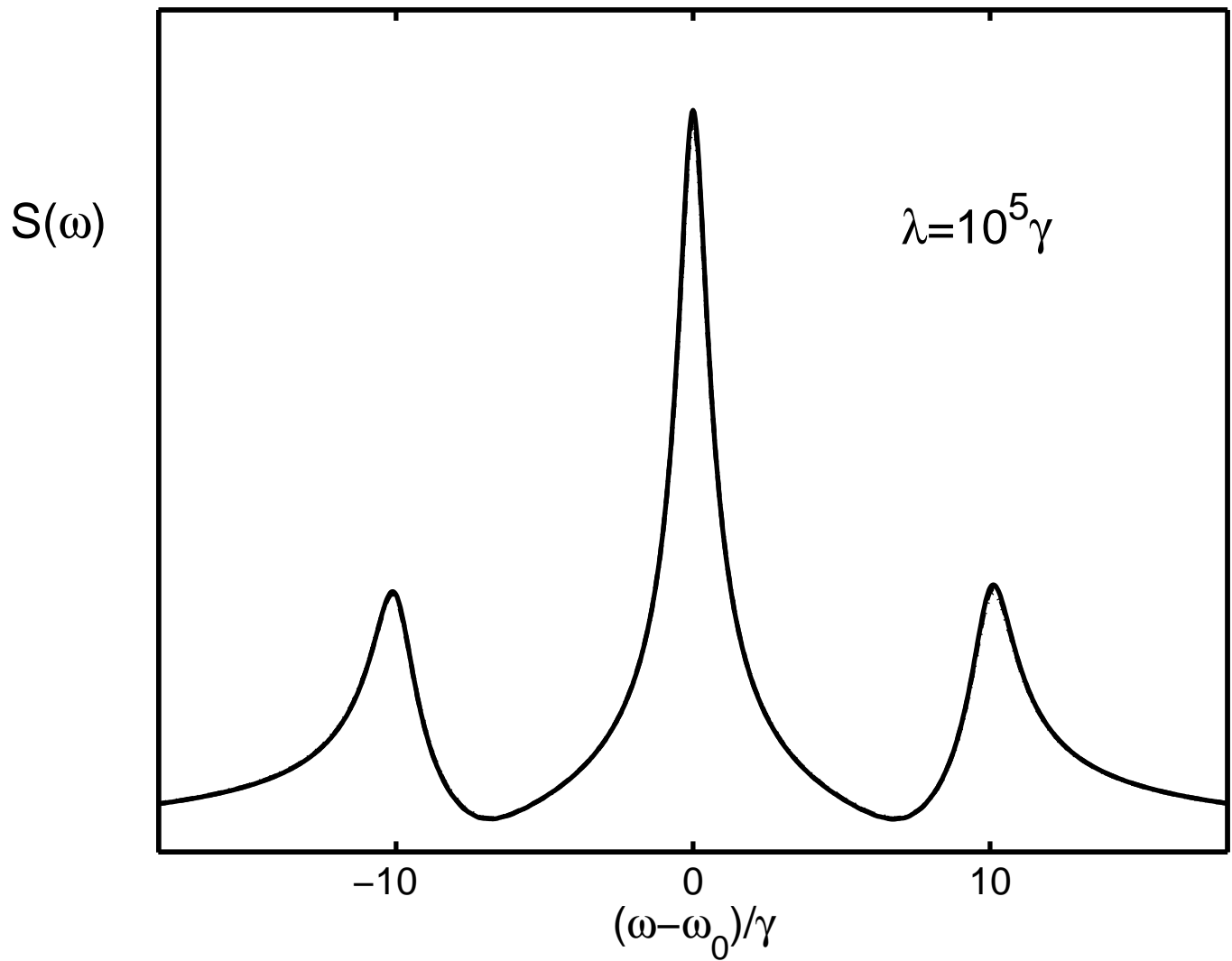
\begin{figure}[tbp]
\caption{Plot of the internal spectrum for $\protect\lambda =10^{5}\protect%
\gamma$ with other parameters the same as Fig.8. The lines corresponding to
different detunings are indistinguishable.}
\label{fig:bandgapspec2}
\end{figure}


\begin{references}
\bibitem{john87}  S. John, Phys. Rev. Lett. {\bf 58}, 2486 (1987).

\bibitem{ho90}  K.~M. Ho, T.~J. Chan, and C.~M. Soukoulis, Phys. Rev. Lett. 
{\bf 65}, 3152 (1990).

\bibitem{joannopoulos97}  J.~D. Joannopoulos, P.~R. Villeueuve, and C.~M.
Soukoulis, Nature (London) {\bf \ 386}, 143 (1997).

\bibitem{yablonovich91}  E. Yablonovich, T.~J. Gmitter, and K.~M. Leung,
Phys. Rev. Lett. {\bf 67}, 2295 (1991).

\bibitem{Gruning95}  U. Gruning, V. Lehmann, and C.~M. Engelhardt, Appl.
Phys. Lett. {\bf 66}, 3254 (1995).

\bibitem{Gruning96}  U. Gruning, V. Lehmann, S. Ottow, and K. Busch, Appl.
Phys. Lett. {\bf 68}, 747 (1996).

\bibitem{Ozbay94}  E. Ozbay {\it et~al.}, Appl. Phys. Lett. {\bf 60}, 2059
(1994).

\bibitem{Soukoulis96}  {\em {NATO} {Advanced Study} Institute Series E:
Applied Sciences}, edited by C.~M. Soukoulis (Kluwer, Dordrecht, 1996),
Vol.~315.

\bibitem{yablonovitch87}  E. Yablonovitch, Phys. Rev. Lett. {\bf 58}, 2059
(1987).

\bibitem{john94}  S. John and T. Quang, Phys. Rev. A {\bf 50}, 1764 (1994).

\bibitem{john84}  S. John, Phys. Rev. Lett. {\bf 53}, 2169 (1984).

\bibitem{john90}  S. John and J. Wang, Phys. Rev. Lett. {\bf 64}, 2418
(1990).

\bibitem{john95}  S. John and T. Quang, Phys. Rev. Lett. {\bf 74}, 3419
(1995).

\bibitem{quang}  T. Quang and S. John, Phys. Rev. A {\bf 56}, 4273 (1997).

\bibitem{molmer99}  K. M{\o}lmer and S. Bay, Phys. Rev. A {\bf 59}, 904
(1999).

\bibitem{hope97}  J.~J. Hope, Phys. Rev. A {\bf 55}, R2531 (1997).

\bibitem{moy97}  G.~M. Moy and C.~M. Savage, Phys. Rev. A {\bf 56}, R1087
(1997).

\bibitem{moy99}  G.~M. Moy, J.~J. Hope, and C.~M. Savage, Phys. Rev. A {\bf %
59}, 667 (1999).

\bibitem{hope00}  J.~J. Hope, G.~M. Moy, M.~J. Collett, and C.~M. Savage,
Phys. Rev. A 023603 (2000).

\bibitem{Hope99}  J.~J. Hope, G.~M. Moy, M.~J. Collett, and C.~M. Savage,
Opt. Comm. {\bf 179}, 571 (2000).

\bibitem{vats}  N. Vats and S. John, Phys. Rev. A {\bf 58, }4168 (1998).

\bibitem{louisell}  W. Louisell, {\em Quantum statistical properties of
radiation} (John Wiley \& Sons, Inc., New York, 1990).

\bibitem{jack00}  M.~W. Jack and M.~J. Collett, Phys. Rev. A {\bf 61},
062106 (2000).

\bibitem{qn}  C.~W. Gardiner, {\em Quantum Noise}, {\em Springer series in
synergetics} (Springer-Verlag, New York, 1991).

\bibitem{feynman}  R.~P. Feynman, Phys. Rev. {\bf 84}, 108 (1951).

\bibitem{carmichael}  H. Carmichael, {\em An Open Systems Approach to
Quantum Optics} (Springer-Verlag, New York, 1991).

\bibitem{tan99}  S. M. Tan, J. Opt. B {\bf 1}, 424 (1999);
http://www.qo.phy.auckland.ac.nz/qo/html/qotoolbox.html.

\bibitem{mollow}  B. R. Mollow, Phys. Rev. {\bf 188,} 1969 (1969).

\bibitem{resfl}  H. Carmichael and D. F. Walls, J. Phys. B {\bf 9}, L43
(1976), J. Phys. B {\bf 9}, 1199 (1976).

\bibitem{numerical}  W. H. Press, S. A. Teukolsky, W. T. Vetterling and B.
P. Flannery, {\em Numerical Recipies in C: The Art of Scientific Computing. }%
2nd ed. (Cambridge University Press 1999).

\bibitem{imamoglu}  A. Imamo\={g}lu, Phys. Rev. A {\bf 50}, 3650 (1994).

\bibitem{garraway}  B.~M. Garraway, Phys. Rev. A {\bf 55}, 2290 (1997).

\bibitem{garraway2}  B. M. Garraway, Phys. Rev. A {\bf 55}, 4636 (1997).

\bibitem{diosi98}  L. Di\`{o}si, N. Gisin, and W.~T. Strunz, Phys. Rev. A 
{\bf 58}, 1699 (1998).

\bibitem{struntz99}  W.~T. Strunz, \ L. Di\`{o}si and N. Gisin, Phys. Rev.
Lett. {\bf 82}, 1801 (1999).

\bibitem{breuer99}  H. P. Breuer, B. Kappler, and F. Petruccione, Phys. Rev.
A {\bf 59, }1633 (1999).

\bibitem{jack99}  M.~W. Jack, M.~J. Collett, and D.~F. Walls, J. Opt. B {\bf %
1}, 452 (1999).

\bibitem{collett85}  C.~W. Gardiner and M.~J. Collett, Phys. Rev. A {\bf 31}
, 3761 (1985).


\end{references}
\end{document}